\documentclass[a4paper,11pt]{article}
\usepackage{a4wide}
\usepackage{amsmath}
\usepackage{graphicx}
\usepackage{pictex}

\title{Test particle motion\\in a gravitational plane wave collision background\vspace{5mm}}
\author{Donato \textsc{Bini}$^{\;\dag\,\S\;}$, Gianluca \textsc{Cruciani}$^{\;\ddag\,\S\;}$
and Andrea \textsc{Lunari}$^{\;\S\;}$\\[5mm]
$^\dag$\hspace{2mm}\textit{Istituto per le Applicazioni della
Matematica C.N.R., I--00161 Rome, Italy}\\[1mm]
$^\S$\hspace{2mm}\textit{ICRA- International Centre for
Relativistic Astrophysics, I--00185 Rome, Italy}\\[1mm]
$^\ddag$\hspace{2mm}\textit{Dept. of Physics, University of
Perugia, I--06100 (Italy)}}

\date{December 2, 2002}

\begin{document}

\maketitle

\begin{flushleft}\small
PACS: 0420C\\
Accepted by {\it Class. Quant. Grav.}\\
on November 29, 2002
\end{flushleft}\normalsize

\begin{abstract}
Test particle geodesic motion is analysed in detail for the background spacetimes of the degenerate
Ferrari-Iba\~nez colliding gravitational wave solutions.
Killing vectors have been used to reduce the equations of motion to a first order system of differential equations which have been integrated numerically. The associated constants of the motion have also been used to match the geodesics as they cross over the boundary between the single plane wave and interaction zones.
\end{abstract}

\section{Introduction}

Among the various theoretical arguments that make the study of gravitational waves one of the best arenas for probing classical and quantum gravity theories, those that involve the strongly nonlinear features of general relativity are very promising. The phenomenology related to the collision between two gravitational waves undoubtedly belongs to this category. However, even if the search for new exact solutions of the Einstein equations describing colliding waves essentially ended in the eighties (see e.g. \cite{griff} for a review), the kinematics of test particle motion in such a background spacetime has never been studied in sufficient detail. 

In the literature it is known that the gravitational wave interaction generates a curvature singularity some time after the instant of collision and at a certain distance from the wavefronts. Moreover there exist degenerate solutions due to Ferrari and Iba\~ nez \cite{Fe-Ib} in some of which  a horizon is created instead. The characteristics of geodesic motion in both types of degenerate solutions are studied in the present paper starting from some preliminary results by Dorca and Verdaguer \cite{do-ve1}. In particular we show  that the horizon and singularity are reached in a finite time by those geodesics which fall into it.

Two nontrivial Killing vectors are used to integrate the geodesic equations, reducing them to a first order system of differential equations which can then be integrated numerically. The results are summarized in a series of appropriately chosen figures typical of the various kinds of behaviour that can occur. Finally some properties of the spacetime associated with the Riemann invariants, the Papapetrou fields and the principal null directions are discussed, emphasizing the aspects arising from the asymmetry of these solutions relative to the coordinates spanning the surface of the wave front.

\section{The special class of Ferrari-Iba\~nez degenerate solutions}

In 1987 Ferrari and Iba\~nez \cite{Fe-Ib} found two degenerate solutions of the Einstein equations that with an appropriate choice of the amplitude parameters associated with their strength can be interpreted as describing the collision of two linearly polarized gravitational plane waves propagating along $z$ with opposite directions. Both these solutions are isometric to the
interior of the Schwarzschild metric but one develops a singularity and the other a coordinate horizon:
\begin{equation}\label{met1}
d\,s^2=(1+\sigma \sin t)^2\,(d\,z^2 -d\,t^2)+\frac{1-\sigma \sin
t}{1+\sigma \sin t}\;d\,x^2+ \cos^2 z\,(1+\sigma \sin
t)^2\,d\,y^2,
\end{equation}
where $\sigma =
\pm 1$ ($\sigma=1$ for the metric with horizon and $\sigma=-1$ for
the metric developing singularity).
The interaction region where this form of the metric is valid (designated as \lq\lq Region I" following \cite{do-ve1}) is represented in the $(t,\, z)$ diagram by an isosceles triangle whose vertex (representing the initial event of collision) can be identified with the origin of the coordinate system; the horizon/singularity is mapped onto the base of the filled isosceles triangle in Fig.~1. The only nonvanishing second order Riemann invariant for both cases is the Kretschmann scalar:
\begin{equation}
K_1=\frac{48}{(1+\sigma \sin t)^6}.
\end{equation}

In order to describe the larger spacetime of which this is only one region, one must introduce the two null coordinates
\begin{equation}
u=(t-z)/2\ ,\ \ v=(t+z)/2\ \ \Longleftrightarrow\ \ t=u+v\ ,\ \ z=v-u
\end{equation}
in terms of which the metric (\ref{met1}) takes the form
\begin{equation}\label{met2}
 \begin{split}
  d\,s^2=&-4\;[1+\sigma \sin(u+v)]^2 d\,u\,d\,v\\
       &+\frac{1-\sigma \sin(u+v)}{1+\sigma \sin(u+v)}\;d\,x^2+\cos^2(u-v)[1+\sigma \sin(u+v)]^2d\,y^2.
 \end{split}
\end{equation}
Following Khan-Penrose \cite{kha-pe}, by proper use of the Heaviside step function $H$, one can easily extend the formula for the metric from the interaction region to the remaining parts of the spacetime representing the single wave zones and the flat spacetime zone before the waves arrive. The interaction region corresponds to the triangular region in the $(u,\,v)$ plane bounded by the lines $u=0$, $v=0$ and $u+v=\pi/2$. One need only make the following substitutions in (\ref{met2}):
\begin{equation}\label{k-p}
u\,\rightarrow\,u\;H(u)\qquad\qquad v\,\rightarrow\,v\;H(v)
\end{equation}
that give rise to the four regions
\begin{equation}
\begin{array}{lll}
\quad u\geq 0\,,\;v\geq 0\,, u+v<\pi/2 \qquad& \mbox{Region I} \qquad& \mbox{Interaction region}\\
\quad 0\leq u < \pi/2 \,,\;v<0 & \mbox{Region II} & \mbox{Single $u$-wave region}\\
\quad u<0\,,\;0\leq v < \pi/2 & \mbox{Region III} & \mbox{Single $v$-wave region}\\
\quad u<0\,,\;v<0 & \mbox{Region IV} & \mbox{Flat space}
\end{array}
\end{equation}
as shown in Fig. 1. In this way the extended metric in general is
$C^0$ (but not $C^1$) along the null boundaries $u=0$ and $v=0$.\\
It is worth noting that any calculation concerning the spacetime
metric  can be appropriately done in $u$ and $v$ coordinates,
while in the case of a test particle freely moving in that same
background spacetime, calculations prove to be easier (especially
in Region I) in the coordinate patch $(t,\,z)$; so a switching
between the two coordinate patches must always be kept in mind in
reading what follows.
\begin{figure}[t]
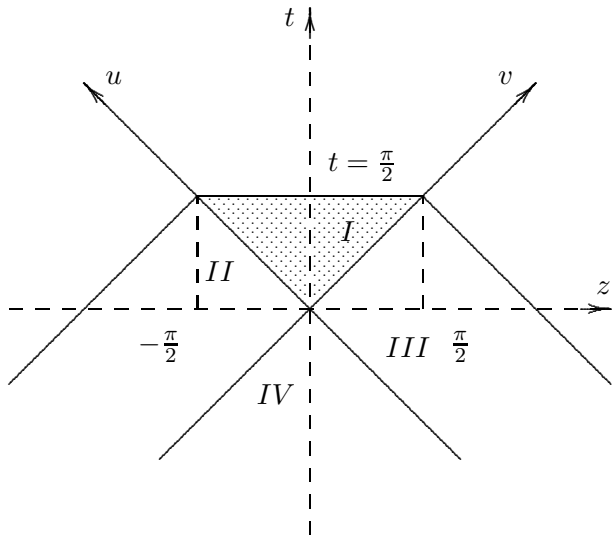

$$
\typeout{pictex fig2}
 \beginpicture
  \setcoordinatesystem units <1cm,1cm> point at 0 0
  \setdashes
   \putrule from  -4   0      to    4  0
   \putrule from   0  -3      to    0  4

\putrule from 1.5 0 to 1.5 1.5 \putrule from -1.5 0 to -1.5 1.5
  \setsolid \setlinear
    \putrule from -1.5 1.5 to 1.5 1.5
    \plot  0  0    3 3    /
    \plot  0  0   -3 3    /
    \plot -2 -2    0 0    /
    \plot  2 -2    0 0    /
    \plot  4 -1  1.5 1.5  /
    \plot -4 -1 -1.5 1.5  /
  \setshadegrid span <.025in>
   \vshade -1.5 1.5 1.5   0 0 1.5   1.5 1.5 1.5 /
  \put {$z$}               [rb]     at      4  0.2
  \put {$t$}               [rt]     at   -0.2  4
  \put {$u$}               [rrb]    at   -2.5  3
  \put {$v$}               [rlb]    at    2.5  3
  \put {$I$}               [rb]     at    0.6  0.9
  \put {$II$}              [rr]     at     -1  0.5
  \put {$III$}             [rl]     at      1 -0.5
  \put {$IV$}              [rt]     at   -0.2 -1

 \put {$t=\frac{\pi}{2}$}             at    0.7   1.9
 \put {$\frac{\pi}{2}$}               at      2  -0.5
 \put {$-\frac{\pi}{2}$}              at     -2  -0.5

  \arrow <.3cm> [.1,.4]    from    3.6    0  to   4 0
  \arrow <.3cm> [.1,.4]    from      0  3.6  to   0 4
  \arrow <.3cm> [.1,.4]    from   2.72 2.72  to   3 3
  \arrow <.3cm> [.1,.4]    from  -2.72 2.72  to  -3 3
\endpicture
$$
\caption{\it The null coordinates and the different regions they induce. The $(t,\,z)$ coordinates are scaled by a factor of $\sqrt{2}$ with respect to the $(u,\, v)$. Region IV is flat space: geodesic paths therein are represented by straight lines whose slope affects the value of the constant $K_v$ in Region II.}\label{fig:1}
\end{figure}

\section{Timelike geodesics}

For both degenerate solution background spacetimes we will solve the geodesic equations
\begin{equation}
\frac{d^2 x^{\alpha}}{d \tau^2}+\Gamma^{\alpha}{}_{\mu\nu}\,( x(\tau) )\; \frac{d x^{\mu}}{d \tau}\;\frac{d x^{\nu}}{d \tau}=0
\end{equation}
with the 4-velocity normalization condition specifying timelike orbits
\begin{equation}
g_{\mu\nu}\,(x(\tau))\;\frac{d x^{\mu}}{d\tau}\;\frac{d x^{\nu}}{d \tau}=-1
\end{equation}
and where the coordinates, metric components and the Christoffel symbols depend on the proper time parameter $\,\tau\,$ along the geodesics as explicitly indicated. Our goal here is to see how a massive particle freely transits from Region IV to Region II (or to its symmetric counterpart, Region III) and finally to Region I, as far as the chosen coordinate patches allow its geodesic motion to be described. Since the spacetime here admits 4 independent Killing vector fields, one can reduce the second order equations of motion to a first order system where three new constants (one for each Killing vector, two of them appearing always in a fixed analytic relation between each other: this leaves three) supplement the normalization condition above. One can then solve these four conditions for the first order derivatives $U^{\mu}=d x^\mu/d \tau $ to obtain this system.

For each Killing vector $\,\xi$ of the metric one obtains a constant of the motion from
\begin{equation}
\xi{}_{\,\mu}\;U^{\,\mu}=K
\end{equation}
along each geodesic. In our case, for both kinds of metric, there are two obvious spacelike translational Killing vectors: $\xi_{(1)}=\partial_x$ and $\xi_{(2)}=\partial_y$, with associated constants $K_x=U\cdot \xi_{(1)}$ and $K_y=U\cdot \xi_{(2)}$, which always exist in both the single-wave and the interaction regions, since the metric (\ref{met1}) there does not depend on $x$ or $y$. For the same reason, in Region II (or Region III) a third translational Killing vector is immediately found from the independence of the metric (\ref{met2}) on $v$ (or on $u$), once expressed using (\ref{k-p}); the associated constant will be designated by $K_v=U\cdot \xi^{\;\mbox{\scriptsize II}}_{(3)}$ (or $K_u=U\cdot \xi^{\;\mbox{\scriptsize III}}_{(3)}$).
It is also easy to see that two additional Killing vectors exist in Region I
\begin{align}
& \xi^{\;\mbox{\scriptsize I}}_{(3)}=\cos y\ \partial_z + \sin y\ \tan z\ \partial_y\\
& \xi^{\;\mbox{\scriptsize I}}_{(4)}=-\sin y\ \partial_z + \cos y\ \tan z\ \partial_y
\end{align}
Only one further constant is needed, which can be taken to be the square root of the sum of their two associated constants of motion
\begin{equation}
K_z=\sqrt{K_{z1}^2+K_{z2}^2}
\end{equation}
where $K_{z1}=U\cdot \xi^{\;\mbox{\scriptsize I}}_{(3)}$ and $K_{z2}=U\cdot \xi^{\;\mbox{\scriptsize I}}_{(4)}$. The timelike geodesics in Region I are then found to satisfy a subsystem of two first-order differential equations for motion in the $t$-$z$ (or $u$-$v$) plane depending on three constants $K_x,\;K_y,\;K_z$. For the sake of clarity, the Killing symmetries used are synthesized in Table 1.

\begin{table}[t]\center
\renewcommand{\arraystretch}{1.3}
\renewcommand{\tabcolsep}{2mm}
\begin{tabular}{|c|c|c|c|c|c|c|c|c|}
\hline
\multicolumn{3}{|c|}{Region II} & \multicolumn{3}{|c|}{Region I} & \multicolumn{3}{|c|}{Region III}\\
\hline
 & {\small Coordinate} & {\small Constant} &  & {\small Coordinate} & {\small Constant} &  & {\small Coordinate} & {\small Constant}\\
\hline
$\xi_{(1)}$ & $x$ & $K_x$ & $\xi_{(1)}$ & $x$ & $K_x$ & $\xi_{(1)}$ & $x$ & $K_x$\\
\hline
$\xi_{(2)}$ & $y$ & $K_y$ & $\xi_{(2)}$ & $y$ & $K_y$ & $\xi_{(2)}$ & $y$ & $K_y$\\
\hline
$\xi^{\;\mbox{\scriptsize II}}_{(3)}$ & $v$ & $K_v$ & $\xi^{\;\mbox{\scriptsize I}}_{(3)}$ & $-$ & $K_{z1}$ & $\xi^{\;\mbox{\scriptsize III}}_{(3)}$ & $u$ & $K_u$\\
\hline
\multicolumn{3}{c|}{} & $\xi^{\;\mbox{\scriptsize I}}_{(4)}$ & $-$ & $K_{z2}$ & \multicolumn{3}{|c}{}\\ \cline{4-6}
\end{tabular}
\caption[c]{\it The Killing vectors in the different regions and the associated coordinates and constants they are referred to in the text.}
\end{table}

Note that by introducing the following differential operator:
\begin{equation}
\mathcal{E}^2=(\xi_{(2)})^2+(\xi^{\;\mbox{\scriptsize I}}_{(3)})^2+(\xi^{\;\mbox{\scriptsize I}}_{(4)})^2
\end{equation}
and by letting it act on a function of the coordinates which oscillates along $(x,\,y)$ with proper frequencies given by $(K_x,\,K_y)$, namely $f(t,\,z,\,x,\,y)=e^{i(K_x\,x +K_y\,y)}\,g(t,\,z)$, it is easy to see that, for any $g(t,\,z)$, the following property is verified:
\begin{equation}
\mathcal{E}^2 (f)= e^{i(K_xx +K_yy)} \left[\frac{\partial^2 g(t,\,z)}{\partial z^2}-\tan z\, \frac{\partial g(t,\,z)}{\partial z}-\frac{K_y^2}{\cos^2 z}\,g(t,\,z)\right]
\end{equation}
which is clearly reminiscent of the square of the angular momentum operator, if one thinks at the specialization of the well-known Chandrasekhar-Xanthopoulos isometry \cite{CX1} to the Ferrari-Iba\~nez degenerate solutions, that relates them to the \lq\lq interior" Schwarzschild spacetime (i.e. for $r<2\,m$ using standard Boyer-Lindquist coordinates). 
Actually it is exactly the operator associated with the integrable part of the Klein-Gordon equation (i.e. the $z$ part),
already studied by Dorca and Verdaguer and Yurtsever \cite{do-ve1,yurt}: this gives
the Killing vectors explicitly presented here a special importance and also helps to explain the integrability of the Klein-Gordon equation itself.

To extend a geodesic from Region II to Region I the value of $K_z$ must be selected properly. 
The geodesic system in Region I is\\
\begin{align}\label{sist3}
& \frac{d t(\tau)}{d\tau}=\frac{1}{(1+\sigma\sin t(\tau))}\sqrt{1+\frac{K_x^2\,(1+\sigma\sin
t(\tau))}{1-\sigma\sin t(\tau)}+\frac{K_y^2+K_z^2}{(1+\sigma\sin t(\tau))^2}}\nonumber\vspace{5mm}\\
& \frac{d z(\tau)}{d\tau}=\frac{1}{(1+\sigma\sin t(\tau))^2}\sqrt{K_z^2-K_y^2\,\tan^2 z(\tau)}\nonumber\vspace{5mm}\\
& \frac{d x(\tau)}{d\tau}=K_x\,\frac{1+\sigma\sin t(\tau)}{1-\sigma\sin (\tau)}\nonumber \vspace{5mm}\\
& \frac{d y(\tau)}{d\tau}=K_y\,\frac{1}{\cos^2 z(\tau)\,(1+\sigma\sin t(\tau))^2}
\end{align}
while in Region II it is\\
\begin{align}\label{sist4}
& \frac{d u(\tau)}{d\tau}=-\frac{K_v}{2}\,\frac{1}{(1+\sigma\sin u(\tau))^2}\nonumber\vspace{5mm}\\
& \frac{d v(\tau)}{d\tau}=-\frac{1}{2\,K_v}\left[1+K_x^2\,\frac{1+\sigma\sin u(\tau)}{1-\sigma\sin u(\tau)}+K_y^2\,\frac{1}{\cos^2 u(\tau)\,(1+\sigma\sin u(\tau))^2}\right]\nonumber\vspace{5mm}\\
& \frac{d x(\tau)}{d\tau}=K_x\,\frac{1+\sigma\sin u(\tau)}{1-\sigma\sin (\tau)}\nonumber \vspace{5mm}\\
& \frac{d y(\tau)}{d\tau}=K_y\,\frac{1}{\cos^2 u(\tau)\,(1+\sigma\sin u(\tau))^2}
\end{align}\\
Limiting the analysis to the $(t, z)$ plane, we see that one of the most important differences between the horizon and singularity cases lies in the behaviour of the geodesics approaching $t=\pi/2$ (or $u+v=\pi/2$).
\begin{itemize}
\item {\it horizon case} ($\sigma=1$)\\
The slope of the trajectories in the limit $t=\pi/2$ is
\begin{equation}
\left.\frac{d\,z}{d\,t}\right|_{\,t\rightarrow\pi/2}=0
\end{equation}
(or $d u/d v =1$) for $K_x\neq 0$, i.e. the test particle crosses the horizon with zero velocity in the direction of propagation of the waves, regardless for the values of $K_y$ and $K_z$ (see Fig. 5). Such behaviour is instead absent in the case $K_x=0$ (see Figs. 3 and 7), where one finds
\begin{equation}
\left.\frac{d\,z}{d\,t}\right|_{\,t\rightarrow
\pi/2}=\sqrt{\frac{K_z^2-K_y^2\,\tan^2 z}{4+K_y^2+K_z^2}}
\end{equation}
clearly depending on the values of $K_y$ and $K_z$.
The fact that making $K_x\neq0$ changes the behaviour of the geodesics at the horizon so much compared to the $K_x=0$ case corresponds to a sort of \lq\lq broken symmetry" between the $x$ and $y$ coordinates associated with the plane symmetry of these spacetimes, which we will briefly discuss in the next section.

\item {\it singularity case} ($\sigma=-1$)\\
Either for $K_x=0$ or $K_x\not =0$, we have: 
\begin{equation}
\left.\frac{d\,z}{d\,t}\right|_{\,t\rightarrow
\pi/2}=\sqrt{\frac{K_z^2-K_y^2\,\tan^2 z}{K_y^2+K_z^2}}
\end{equation}
and the test particle approaches the singularity with $z$--velocity in general
different from 0. However, here it is important to distinguish whether $K_y$ vanishes or not. For $K_y=0$ the geodesics exhibit a typical twofold behaviour (see Figs. 2 and 4) depending on whether they enter the interaction region or not. This is due to the existence of a critical value of $K_v$, here denoted by $K_v^{\rm thr}$, in the single-wave region, or equivalently to a threshold for the velocity along the $z$ axis in the vacuum region. Geodesics in Region II with
$K_v=K_v^{\rm thr}$ can be extended into Region I with $K_z=0$, i.e. they remain at a fixed $z$ position, irrespective of the presence of the waves; this $K_v^{\rm thr}$ and the Region II-Region I crossing value $u^*$ of $u$ are related by:
\begin{equation}
K_v^{\rm thr}=(\sin u^*-1)\sqrt{1+K_x^2\,\frac{1-\sin u^*}{1+\sin
u^*}}\;.
\end{equation}
For particles exceeding this value of $K_v$, the geodesics are of Type I, i.e. entering the interaction region and reaching the singularity with $d z/d t\rightarrow 1$, otherwise, they are of Type II, i.e. confined in the single-wave region and approaching the singularity with $d z/d t \rightarrow -1$.

As stated above, along the geodesic corresponding to the critical value of $K_v$, the particle enters
at a certain position the interaction region with $d z/d t=0$ and doesn't move. This geodesic marks a triangular part of Region I, also delimited by the singularity and the $u$ axis, which actually doesn't allow any geodesic trajectory to enter.

For $K_y\not =0$ the threshold disappears, as shown in Fig. 6.
\end{itemize}

As for the proper times of those geodesics entering the interaction region and approaching the line $t=\pi/2$, either in the horizon or singularity cases, the situation is summarized in Tables 1  and 2,
for the case of a particle entering Region II at $v=-\pi/2$, with arbitrary choices of $K_x,\,K_y$ and giving $K_v$ the value that corresponds to zero initial velocity along $z$ (i.e., $d u/d v = 1$ for $u \rightarrow 0$), which is $\,-\sqrt{K_x^2+K_y^2+1}\,$.

\begin{table}[t]\center
\renewcommand{\arraystretch}{1.3}
\renewcommand{\tabcolsep}{2.1mm}
\begin{tabular}{|c|c|c|c|c|c||c|c|c|c|c|c|}
\hline $K_x$ & $K_y$ & $-K_v$ & $\tau_{\;\mbox{\scriptsize
II}\rightarrow \mbox{\scriptsize I}}$ & $\tau_{\;\mbox{\scriptsize
I}\rightarrow \mbox{\scriptsize hor}}$ &
$\tau_{\;\mbox{\scriptsize tot}}$ & $K_x$ & $K_y$ & $-K_v$ &
$\tau_{\;\mbox{\scriptsize II}\rightarrow \mbox{\scriptsize I}}$ &
$\tau_{\;\mbox{\scriptsize I}\rightarrow \mbox{\scriptsize hor}}$
& $\tau_{\;\mbox{\scriptsize tot}}$
\\ \hline \hline 0 & 0 & $1$ & 3.1416 & 1.2975 & 4.4391 & 2 & 0 & $\sqrt{5}$ & 0.7874 & 0.2026 & 0.9900\\
\hline 0 & 1 & $\sqrt{2}$ & 2.6492 & 0.8703 & 3.5195 & 2 & 1 & $\sqrt{6}$ & 0.7730 & 0.1920 & 0.9650\\
\hline 0 & 2 & $\sqrt{5}$ & 1.8565 & 0.4883 & 2.3448 & 2 & 2 & $3$ & 0.7348 & 0.1686 & 0.9034\\
\hline 0 & 3 & $\sqrt{10}$ & 1.3545 & 0.3256 & 1.6801 & 2 & 3 & $\sqrt{14}$ & 0.6835 & 0.1442 & 0.8278\\
\hline 1 & 0 & $\sqrt{2}$ & 1.4140 & 0.3585 & 1.7725 & 3 & 0 & $\sqrt{10}$ & 0.5378 & 0.1391 & 0.6769\\
\hline 1 & 1 & $\sqrt{3}$ & 1.3447 & 0.3129 & 1.6576 & 3 & 1 & $\sqrt{11}$ & 0.5330 & 0.1354 & 0.6684\\
\hline 1 & 2 & $\sqrt{6}$ & 1.1846 & 0.2406 & 1.4251 & 3 & 2 & $\sqrt{14}$ & 0.5194 & 0.1260 & 0.6454\\
\hline 1 & 3 & $\sqrt{11}$ & 1.0124 & 0.1874 & 1.1998 & 3 & 3 & $\sqrt{19}$ & 0.4993 & 0.1142 & 0.6135\\
\hline
\end{tabular}
\caption[c]{\it (Horizon case) the proper times for different initial data sets.} \vspace{6mm}
\begin{tabular}{|c|c|c|c|c|c||c|c|c|c|c|c|}
\hline $K_x$ & $K_y$ & $-K_v$ & $\tau_{\;\mbox{\scriptsize
II}\rightarrow \mbox{\scriptsize I}}$ & $\tau_{\;\mbox{\scriptsize
I}\rightarrow \mbox{\scriptsize sing}}$ &
$\tau_{\;\mbox{\scriptsize tot}}$ & $K_x$ & $K_y$ & $-K_v$ &
$\tau_{\;\mbox{\scriptsize II}\rightarrow \mbox{\scriptsize I}}$ &
$\tau_{\;\mbox{\scriptsize I}\rightarrow \mbox{\scriptsize sing}}$
& $\tau_{\;\mbox{\scriptsize tot}}$
\\ \hline \hline 0 & 0 & $1$ & - & - & - & 2 & 0 & $\sqrt{5}$ & - & - & - \\
\hline 0 & 1 & $\sqrt{2}$ & 0.5034 & 0.0001 & 0.5035 & 2 & 1 & $\sqrt{6}$ & 0.2908 & $<10^{-4}$ & 0.2908 \\
\hline 0 & 2 & $\sqrt{5}$ & 0.3174 & 0.0003 & 0.3177 & 2 & 2 & $3$ & 0.2374 & $<10^{-4}$ & 0.2374 \\
\hline 0 & 3 & $\sqrt{10}$ & 0.2240 & 0.0003 & 0.2243 & 2 & 3 & $\sqrt{14}$ & 0.1901 & 0.0001 & 0.1902 \\
\hline 1 & 0 & $\sqrt{2}$ & - & - & - & 3 & 0 & $\sqrt{10}$ & - & - & - \\
\hline 1 & 1 & $\sqrt{3}$ & 0.4112 & $<10^{-4}$ & 0.4112 & 3 & 1 & $\sqrt{11}$ & 0.2148 & $<10^{-4}$ & 0.2148 \\
\hline 1 & 2 & $\sqrt{6}$ & 0.2903 & 0.0001 & 0.2904 & 3 & 2 & $\sqrt{14}$ & 0.1904 & $<10^{-4}$ & 0.1904 \\
\hline 1 & 3 & $\sqrt{11}$ & 0.2139 & 0.0002 & 0.2141 & 3 & 3 & $\sqrt{19}$ & 0.1634 & $<10^{-4}$ & 0.1634 \\
\hline
\end{tabular}
\caption[c]{\it (Singularity case) the proper times for different initial data sets. (-): the geodesic doesn't reach Region I. It is evident that, while the particles approach the horizon smoothly, the formation of a singularity makes their transit in Region I almost immediate.}
\end{table}

\section{Papapetrou fields and principal null directions in Region I}

The Killing vector field $\xi_{(1)}$ generates a Papapetrou field $F$ \cite{fa-so} whose principal null directions are aligned with those (repeated) of the spacetime itself (which is of Petrov type D). The key properties are the following:
\begin{align}
& E(X)_{\alpha \mu}\equiv C_{\alpha\beta\mu\nu}\,X^\beta\,X^\nu =
-3\,(1-\sigma \sin t)\,[a(X)\otimes a(X)]^{\rm (TF)}\;,\nonumber\\
& H(X)_{\alpha \mu}\equiv
{}^*C_{\alpha\beta\mu\nu}\,X^\beta\,X^\nu=0
\end{align}
where $X=\xi_{(1)}/||\xi_{(1)}||\equiv \sqrt{g_{xx}}\;d\,x$, and
\begin{equation}
a(X)=\nabla_X X \equiv - [\partial_\mu \ln
\sqrt{g_{xx}}\,]\;d\,x^\mu = \frac{\sigma}{\cos t}\;d\,t
\end{equation}
is the (timelike) acceleration field associated to $X$ and ${\rm
(TF)}$ stands for the trace free part of a tensor. Explicitly, for
a generic $2 \choose  2$ tensor  A, orthogonal to $X$, one has:
\begin{equation}
A^{\rm (TF)}_{\alpha\beta}= A_{\alpha\beta} -\frac13\,[{\rm
Tr}A]\,P(X)_{\alpha\beta}\;,\qquad P(X)=g-X\otimes X\;,
\end{equation}
with $P(X)$ being the projector orthogonal to the spacelike $X$
direction. Finally the Papapetrou field associated to $\xi_{(1)}$
is
\begin{equation}
F_{\alpha\beta}=\xi_{(1)}{}_{\alpha ;\beta} = [\xi_{(1)}\wedge
a(X)]_{\alpha\beta}\;.
\end{equation}
The two independent principal null directions of both
$F_{\alpha\beta}$ and $C_{\alpha\beta\gamma\delta}$ are
\begin{align}
& l=\frac{1}{2\,\cos t} \left[ \partial_t +\frac{(1+\sigma \sin t)^2}{\cos t}\, \partial_x \right]\nonumber\\
& n=\frac{\cos t}{(1+\sigma \sin t)^2}\;\partial_t -\partial_x\ ,
\end{align}
affinely parametrized and normalized so that $l\cdot n =-1$. From these expressions it is clear that the special role of the Killing vector $\xi_{(1)}$ (compared to  $\xi_{(2)}$) is due to the fact that it belongs to the $2$-plane spanned by the principal null directions $l$ and $n$. This, at least from a purely geometrical point of view, leads to a difference between the $x$ and $y$ coordinates in terms of which the metric is expressed.

\section{Concluding remarks}
A detailed analysis of (timelike) geodesic motion has been carried out in the spacetimes of the
Ferrari-Iba\~ nez degenerate solutions representing two colliding gravitational plane waves. The study of these simple cases is a useful exercise in general relativity because they contain the main features related to the nonlinear collision of waves, i.e., the creation of a horizon or that of a curvature singularity.

\newpage

\begin{figure}[t]
\includegraphics[width=150mm,height=75mm]{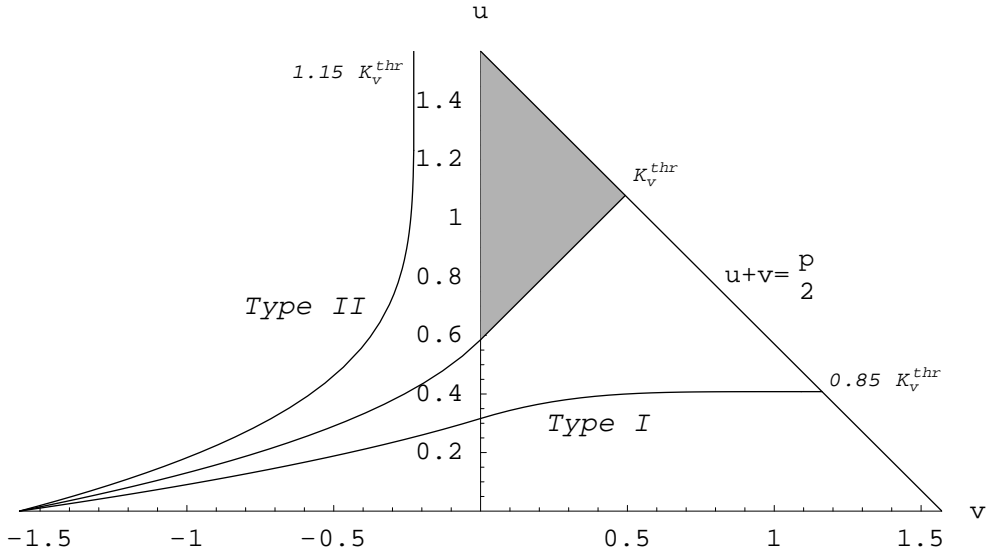}
\caption{For $\sigma=-1$ (metric
with singularity) and $(K_x=0,\,K_y=0)$, the critical initial
value of $K_v$ defined in the text is $K_v^{\rm thr}=-0.447645$. The forbidden
region for timelike geodesics is represented in gray.}
\end{figure}
\begin{figure}[b]
\includegraphics[width=150mm,height=75mm]{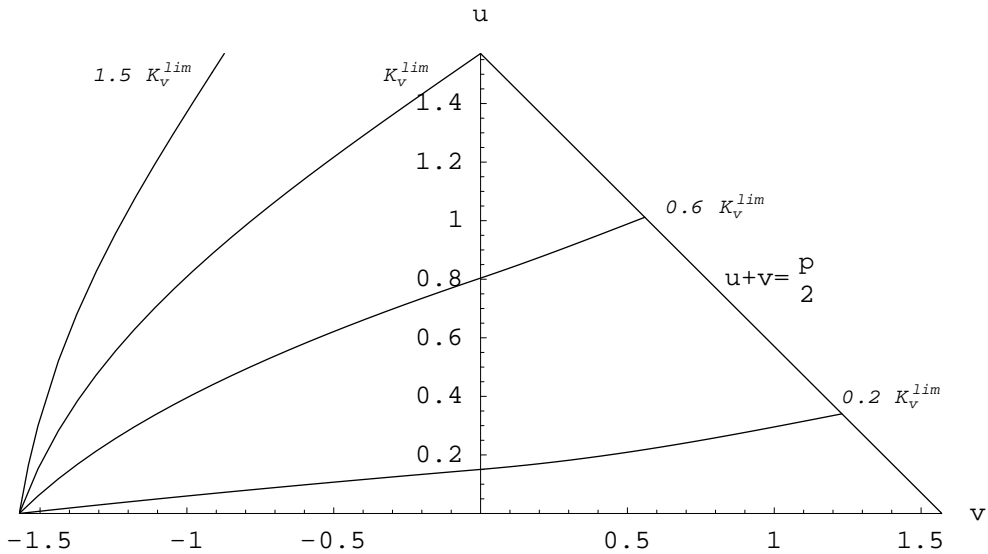}
\caption{For
$\sigma=1$ (metric with horizon) and $(K_x=0,\,K_y=0)$, the
geodesics regularly fill the entire Region I, coming from a set of
initial data characterized by a limiting value of $K_v$, $K_{v}^{\rm lim}=-1.6653$ under
which the particles possess a too large momentum along $z$
(moving into reverse with respect to the single wave) to enter the collision region.}
\end{figure}

\begin{figure}[t]
\includegraphics[width=150mm,height=75mm]{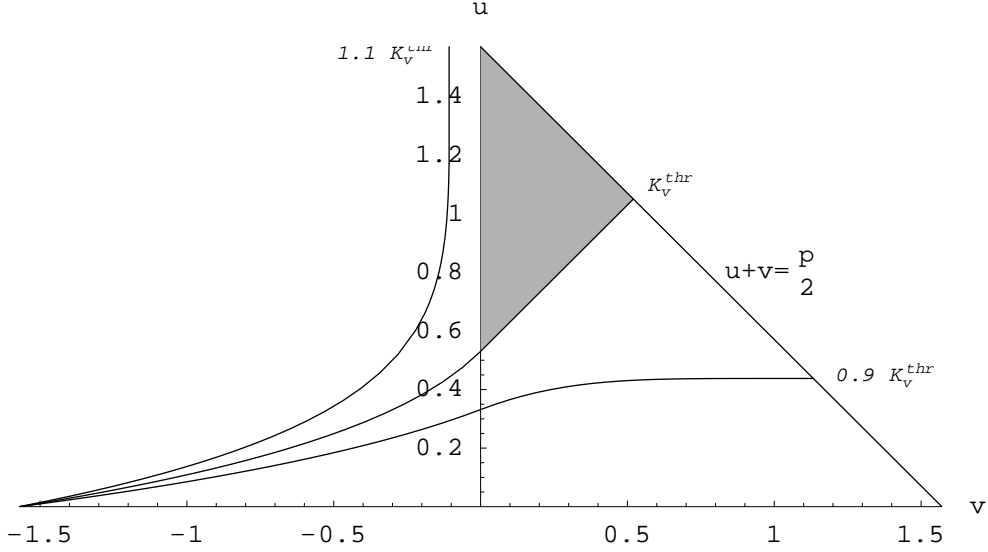}
\caption{For $\sigma=-1$ and
$(K_x=1,\,K_y=0)$, the geodesics qualitatively behave as in the
$K_x=0$ case, with now $K_v^{\rm thr}=-0.569647$; increasing
$K_x$ corresponds to a lowering of the critical geodesic in Region
II (i.e. an enlargement of the \lq\lq forbidden region").}
\end{figure}
\begin{figure}[b]
\includegraphics[width=150mm,height=75mm]{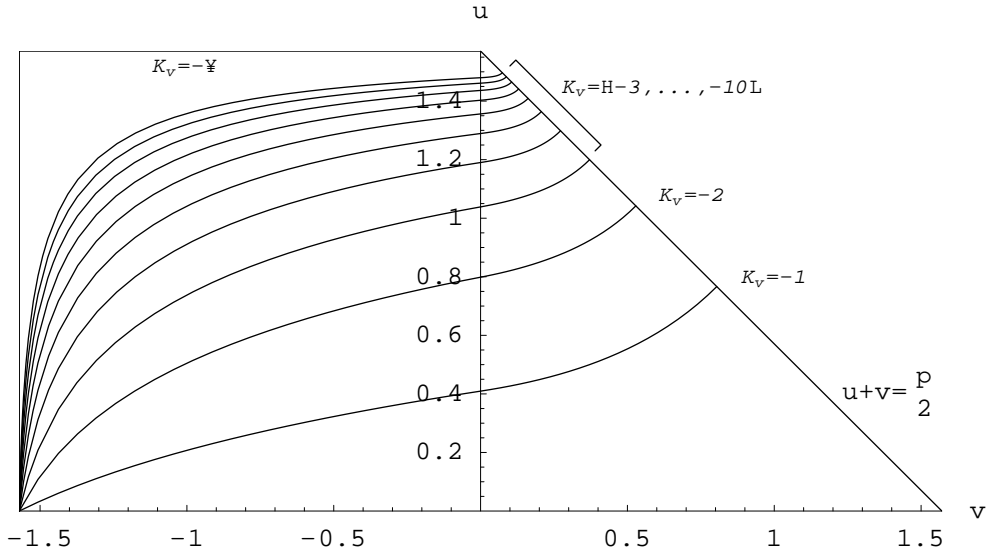}
\caption{For $\sigma=1$ and
$(K_x=1,\,K_y=0)$, the geodesic motion has a global homogeneity
for all values of $K_v$ (it is always the case when $K_x\neq 0$).
As one expects, in Region II the particles are accelerated (in the sense 
that they exhibit a relative acceleration $d^2 z/d t^2$) in the
positive $z$ direction, while in Region I they are decelerated
until they reach zero velocity exactly at the horizon. The limit
$K_v\rightarrow -\infty $ reconstitutes the so-called {\it fold
singularities} already observed in the case of null geodesics by
Dorca and Verdaguer \cite{do-ve1}}
\end{figure}

\begin{figure}[t]
\includegraphics[width=150mm,height=75mm]{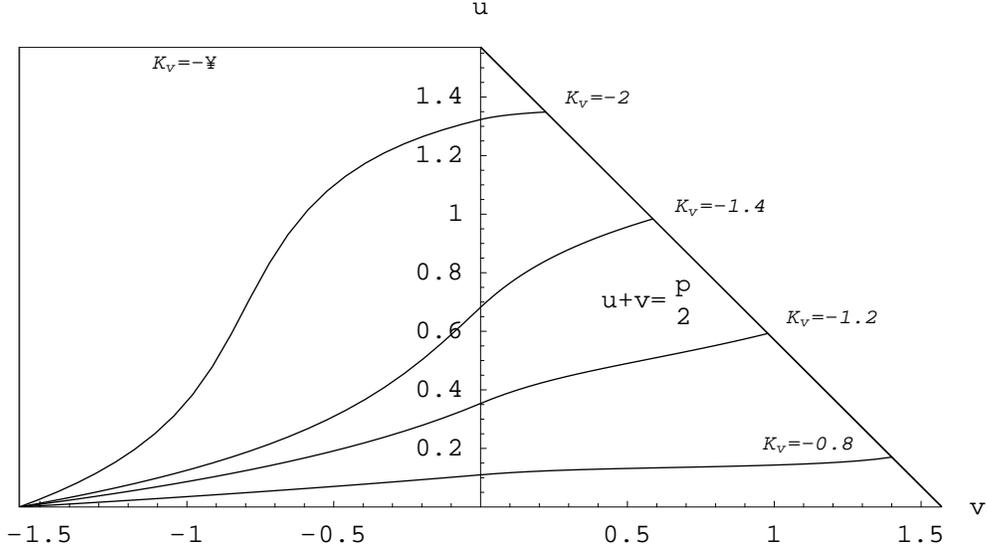}
\caption{The case
$\sigma=-1$ and $(K_x=3,\,K_y=1)$ is a good example of what the
geodesics of the metric with singularity look like when $K_y\neq
0$. The {\it fold singularities} arise for $K_v\rightarrow
-\infty$ while the particles accelerate abruptly (relative acceleration of course) and with a change
in sign along $z$ in Region II. There are no forbidden regions in
Region I.}
\end{figure}
\begin{figure}[b]
\includegraphics[width=150mm,height=75mm]{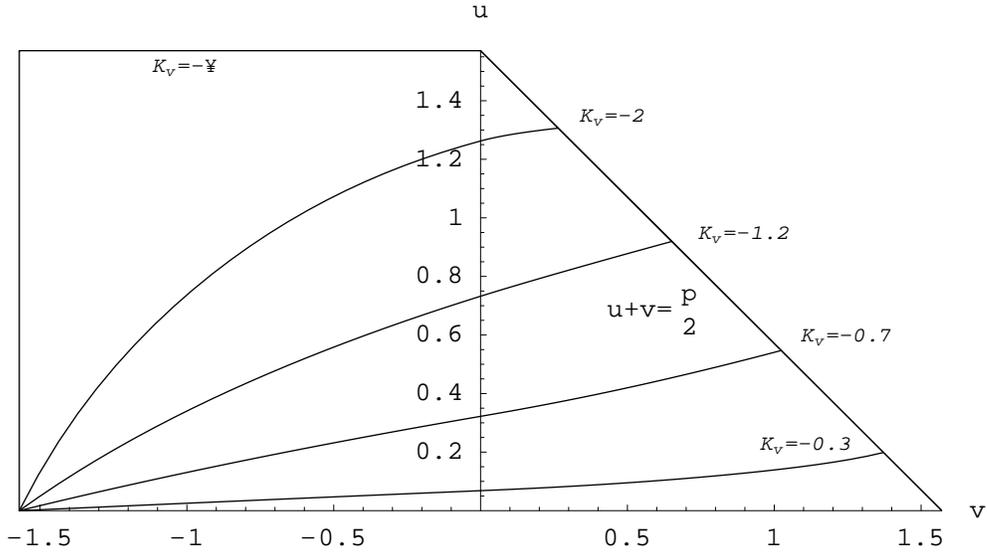}
\caption{For
$\sigma=1$ and $(K_x=0,\,K_y=1)$ (this case qualitatively
characterizes all the sets of initial data with $K_x=0$), the
metric with horizon exhibits a very homogeneous geodesic motion in
Region II, where all the possible values of the momentum along
$z$ are allowed to enter Region I (so generating again the {\it
fold singularities}) and slight changes in Region I, where the
particles decelerate for high entering values of $d z/d t$ and
accelerate otherwise.}
\end{figure}

\end{document}